\documentstyle[aps,prl,preprint,psfig,amsfonts,amssymb]{revtex}
\psfigurepath{../figures}                       % this is toerase without
\def\be{\begin{equation}}
\def\ee{\end{equation}}
\def\beq{\begin{eqnarray}}
\def\eeq{\end{eqnarray}}

\def\inte{\int\!\!}

\def\a{{\mathfrak a}}
\def\A{{\mathfrak A}}
\def\aa{\overline{{\mathfrak a}}}
\def\AA{\overline{{\mathfrak A}}}
\def\e(#1){{\rm e}^{#1}}

\begin{document}
\input{epsf.tex}

\draft

\title{On the problem of $Yb_{4}As_{3}$.  
 }

\author{M S. Laad and J.R. Reyes Martinez$^1$ }
\address{$^1$Institut f\"ur Theoretische Physik, Universit\"at
zu K\"oln,
Z\"ulpicher Str 77, 50937 K\"oln, Germany}

\date{}

\maketitle

\begin{abstract}
  Based on a perusal of available experimental data, we propose
a possible 
explanation for the heavy-fermion Fermi liquid state coexisting
with quasi-one dimensional antiferromagnetism in the
low-carrier pnictide
$Yb_{4}As_{3}$.  Both of these low temperature features are
shown to 
originate from one single fact - the charge ordered state
forming at high-$T$
is imperfect, due to quantum fluctuations in the CO state.  The
quasi-1D 
magnetism is understood in terms of the physics of the squeezed
Heisenberg 
chain, while the heavy fermion-like behavior arises from the
compensation of 
the fraction of the $Yb$ moments (which are disordered) by the
$2p$ band of
$As$.  The proposed picture shows how the seemingly strange
features observed
experimentally can be reconciled as a result of this single
hypothesis. 
       
\end{abstract}
     

\newpage

{\bf 1. INTRODUCTION}

  Rare-earth pnictides like $Yb_{4}As_{3}$ have been
increasingly studied
in the past few years, as they are thought of as being
"low-carrier density
Kondo systems".  Transport measurements show similarities with
typical 
heavy fermion Fermi liquid metals [1], while the thermodynamics
and the 
magnetic response [2] is characteristic of one-dimensional spin
systems.
A naive estimation of the carrier number from the Hall data
leads one to 
conclude that the carrier density is very low [3].  How such a
supposedly 
small carrier density can screen the $Yb$ local moments to give
a heavy FL
at low-$T$ is the issue central to the understanding of this
material.
Purely on theoretical grounds, one might want to ask:    

(1) Under what conditions can a heavy-fermion like FL metal
coexist with 
quasi-1D spin fluctuations?  How do these degrees of freedom
affect each other
at low-$T$?

(2) A more general, but related question:  Are there hitherto
unknown routes to
heavy fermion behavior, other than the well-studied Kondo
effect?

  In this letter, we want to propose a route which ties
together the various
seemingly irrevocable facts that are known about this material.
We start with
a brief recapitulation of known experimental and theoretical
facts, and then
go on to show how both the heavy-fermion like transport and the
lower-
dimensional magnetism arise naturally from {\it imperfect}
nature of the
charge ordering that occurs at higher $T$.  Our analysis also
provides a 
natural explanation of the fact that the closely related
materials $Yb_{4}X_{3}, X=P, Sb$, behave very differently from
$Yb_{4}As_{3}$.  The phosphide is an 
insulator all the way down to the lowest $T$, with a more
perfect charge order, while the antimonide is a 
mixed-valence metal (no charge ordering) down to the lowest
$T$.

  $Yb_{4}As_{3}$ is a semiconductor at room temperature, with
the rare-earth
$f$-ion of $Yb$ in a mixed valence state ($Yb^{14+}-Yb^{13+}$).
At about $293$ K,
 a collective JT distortion causes three of the four identical
diagonals along 
the cubic cell to elongate, leading to a contraction of the
fourth diagonal,
into which the $Yb^{13+}$ ions gather.  Since $Yb^{13+}$
carries a spin $1/2$,
one has an (almost half-filled) chain with strong electronic
correlations.
These are precisely the degrees of freedom giving rise to the
$1D$ spin fluctuations observed in inelastic neutron
scattering, and to the spin susceptibility 
 and the $\gamma$ coefficient of the specific heat, as shown
on [2], and discussed on experimental [2] and theoretical
grounds [3].

  Since the spin degrees of freedom are those of the $1D$
Hubbard model, the
spin excitations are fermionic spinons rather than conventional
bosonic spin
waves.  But this is not all, as bandstructure calculations
reveal that the  
carriers seem to come predominantly from the $As$ 2p-band.  A
consistent picture
 of the low-$T$ transport must necessarily involve the
scattering of these 
carriers on the spin fluctuations of the Hubbard chain.  
If the chains were half-filled, leaving the $As$ p carriers out
of the reasoning would lead one to conclude that the material
must be a Mott-Hubbard insulator.
On the other hand, if one does include the $As$ carriers, a
heavy fermion-like
resistivity can only result from the scattering of these light
carriers off
almost localized $Yb$ 4f-spins.  The issue then is 
whether this scattering (details have not been worked out) can
give rise to a 
quadratic (in $T$) scattering rate.  The large $A$ value
suggests  
strong scattering, but in view of the fact that one is dealing
with spinon-pair
fluctuations that scatter the light carriers, a concrete
calculation is needed
to answer the question.  We argue, however, (see below) that
this mechanism 
is {\it not} responsible for the observed resistivity, which,
we show, has a
more conventional interpretation. 

  This is intimately related to the second question posed
above:  the scattering giving rise to a HF-FL metal at low-$T$
must necessarily involve quenching the
local moment degree of freedom of the Hubbard chain.  So one
might want 
to address the
question of a Kondo-like effect originating in a way different
from the usual
accepted one.  The basic question should then be rephrased as:

Can light electron-like carriers scattering on spinon-pair
fluctuations yield
heavy fermion like behavior?  

or,

Is the above view incorrect, and could a conventional Kondo
effect occur as a 
result of scattering of the $p$-holes off the spin fluctuations
of the fraction 
 of the localized spins in the other three chains?

If the second point of view indeed holds, a consistent
explanation for the 
above anomalous features can indeed be given without having to
take recourse to
exotic scenarios.

  To be able to decide the issue, one has to appeal to
experiment.  The
experimental results need to be correlated and a common thread
connecting the
observed phenomena needs to be searched.  We summarize the
experimental results
below.

(1)  The high-$T$ phase is a mixed-valent metal.  This
undergoes a first-order
charge-ordering transition at $T_{c}=293$K via a collective
band-JT effect.
The effect is to shorten one of the four equivalent chains (at
high $T$) and
to elongate the other three.  The $f$-holes carrying a spin
$S=1/2$ sit on
the shorter chains, and because the Hubbard $U_{ff}>>t_{f}$,
the effective 
$f-f$ hopping strength, one has to deal with a strongly
correlated 1D Hubbard
like chain.  But this is not the whole story, because
bandstructure calculations
 have shown that the carriers responsible for conduction are in
fact derived 
from the $As$ 2p-band (which is a wide band, giving rise to
small effective-mass
 carriers).  The low-$T$ heavy fermion behavior has to do with
the scattering
of the light $p$-holes off the almost localized $f$-moments of
the $Yb$.  The
question for theory is:  How can this give rise to a
resistivity, $\rho(T) \simeq AT^{2}$ with large $A$?

(2)  The specific heat shows a {\it linear} $T$-dependence, and
one might 
naively associate this as one of the signatures of usual
heavy-fermion behavior
especially since the ratio $A/\gamma^{2}$ is also
characteristic of heavy 
fermion metals.  What gives the game away is that a related
material, $Yb_{4}P_{3}$, an insulator, also has the linear
term, with a $\gamma$-coefficient larger
than that for $Yb_{4}As_{3}$!  Thus, the linear specific heat
can only result
from the spin degrees of freedom, spinons, of the 1D Hubbard
model.
That this
is indeed the case is shown conclusively by inelastic neutron
scattering data,
which reveals a spin dispersion which is almost that of an
ideal 1D
$S=1/2$ AFM chain.     
An external magnetic field has been shown to result in
significant
changes in
the $\gamma$ coefficient, as well as in the spin
susceptibility, while leaving 
the resistivity almost unchanged!  Exactly the opposite happens
with the effect
of external pressure, as we discuss below.

  First, we summarize the effect of an external magnetic field.
A uniform 
longitudinal field has been shown to generate a staggered
transverse field
component along the $x$-direction, resulting in the effective
model
given by

\be
H_{ch}=2I\sum_{i}{\bf S_{i}}.{\bf S_{i+1}} -
h\sum_{i}(-1)^{i}S_{i}^{x} - H\sum_{i}S_{i}^{z} 
\ee
Calculation of the specific heat, $C(T)$, shows the opening up
of a gap at low
$T$ in good agreement with experimental observations [4].  The
static 
magnetic susceptibility depends on the orientation of the
external field  
relative to the spin chain.  What is so surprising
 is that this change in the spin excitation spectrum has very
little influence 
on the low-$T$ resistivity.  This is hard to understand if the
1D spin 
excitations are assumed to drive the observed Fermi liquid
resistivity.

  Application of hydrostatic pressure has exactly the reverse
effect:  the 
resistivity changes drastically as pressure is increased, the
main features 
being the suppression of the high-$T$ "charge ordering"
transition, a gradual
but strong reduction of the broad maximum in $\rho(T)$, and a
drastic decrease 
in the coefficient $A$.  The suppression of the high-$T$ CO
state can be 
understood in terms of the increased hybridization between the
As $p$ and the
Yb $4f$ orbitals.  The specific heat and spin susceptibility,
however, are
completely unaffected by pressure. 

  Replacement of $As$ by an ion like $P$ should stabilize the
charge 
ordered (CO) state.  This is indeed observed [5], but with a
larger $\gamma$
coefficient of the specific heat than in $Yb_{4}As_{3}$.  The
resistivity in
this case is {\it insulating} all the way down to low $T$,
confirming that the
HF-FL behavior in $Yb_{4}As_{3}$ is {\it not} related to
scattering off the
1D spinons, but is connected with the imperfect CO state in
$Yb_{4}As_{3}$.
Experimental estimates [6] point to the disordering of the CO
state
with a small fraction of the $Yb^{3+}$ on the three longer
chains.  This observation is 
further supported by LDA+U calculations [6], which show that
the hopping integral {\it between} the short chain and the
longer chains is {\it greater} than the intrachain hopping
integral in the short chain.  These $S=1/2$ spins are
randomly distributed in the three-dimensional host, and scatter
the light $As$
holes.  

  In this paper, we propose that both, (i) the one-dimensional
magnetism  
, and (ii) the low-$T$ heavy-fermion FL metallic phase observed
in $Yb_{4}As_{3}$ result from one single phenomenon: the
incomplete nature of the the CO state,
which sets in as a first-order transition at $T \simeq 293$K
[1].   
This high-$T$ CO state has been sought to be explained by a
collective band
Jahn-Teller distortion [7,8], whereby the homogeneous
mixed-valent state with
non-integral $Yb$ valence state of $2.25$ is unstable to a
state where the 
$Yb^{3+}$ ($S=1/2$) congregate predominantly on one chain.
This chain is 
shortened, while the other three are elongated.  However, this
CO is not perfect
 and estimates as well as an explicit calculation of a Hubbard
model with the
Jahn-Teller distortions in the geometry of four
interpenetrating chains
relevant to $Yb_{4}As_{3}$ 
[6] show that a small fraction of the $Yb^{3+}$ sites are on
the other chains, an observation that is reconcilable with the
fact that the 
hopping integral between the short and the long chains is
greater than that 
along the short chain.  This observation is crucial for the
picture we propose
below.  

  The 1D magnetism observed in this material is then understood
in terms of the
physics of the 1D Hubbard (or squeezed Heisenberg model [9])
chain.  As is 
known [10], the spectrum of the 1D Hubbard model is
characterized by spin-charge
 separation.  The spin is carried by neutral fermionic
excitations, i.e, by the
spinons, while the charge degrees of freedom are described by
spinless, hard 
core bosons, the holons.  Because of spin-charge separation,
one can continue 
using the Heisenberg model (with an exchange coupling that
depends on the 
deviation from half-filling) to describe the spin dynamics
[11].    

  We have computed the uniform static spin susceptibility and
the specific
heat from the thermodynamic Bethe ansatz (TBA) equations [12]
for a modified 
Heisenberg chain with an exchange strength (dependent on the
hole concentration)
 given by 
\be
I_{eff}(n)= n[1 -\sin( 2\pi n)/2 \pi n]
\ee

  We point out that such an approach has also been used by Zvyagin [17]
in the context of $Yb_{4}As_{3}$.  However, while we agree with [17] on the
issue of the quasi-one dimensional magnetism in $Yb_{4}As_{3}$, we differ
on the origin of the FL-like resistivity (see below).
The Hamiltonian is,

\be
H_{ch}= 2I_{eff}(n)\sum_{i}{\bf S_{i}}.{\bf S_{i+1}}
\ee
  It is well known that the 1D Heisenberg spin model is
completely integrable,
and that the TBA gives an exact expression for the free energy
of the system.
Using the spinon formulation of the TBA (using the quantum
transfer
matrix) [12] 
 the problem reduces to the solution of two coupled nonlinear
integral
 equations.  
\beq
 \label{a(x)}
\ln\a(x) &=& -\frac{\pi I_{eff}}{T\cosh x}
+(k\ast\ln\A)(x) - (k\ast\ln\AA)(x- i\pi
+i\epsilon),\\\nonumber 
\ln\aa(x) &=& -\frac{\pi I_{eff}}{T\cosh x}
+(k\ast\ln\AA)(x) - (k\ast\ln\A)(x + i\pi
-i\epsilon),\\\nonumber 
 \A(x) &=& 1 + \a(x),\qquad \AA(x) = 1 +\aa(x).
\eeq
\begin{equation}
k(x) :=
\frac{1}{2\pi}\int_{0}^{\infty}\!\!
\frac{\e(-\frac{\pi}{2}k)}{2\cosh\frac{\pi}{2}k}
\cos(kx)\hbox{d}k,
\end{equation}
where the asterik denotes convolution $(f\ast g)(x) :=
\int_{-\infty}^{\infty}\!\! f(x-y)g(y)\hbox{d}x$.  The free
energy is
given by $f(0)$ with
\begin{equation}
\label{f}
f(x) = -T\inte\frac{\ln[\A(y)\AA(y)]}{\cosh(x - y)}\hbox{d}y,
\end{equation}
ignoring the contribution of the groundstate energy.
Solving these equations by the fast-fourier transform method,
we have 
computed the spin susceptibility and the specific heat as a
function of $T$ for
the case applicable to $Yb_{4}As_{3}$, i.e, with a small
fraction of spins in the
other three chains ($n=0.9$).  Calculation of the staggered
spin susceptibility
has already been carried out in [4], and is consistent with a
gapped spin-wave 
spectrum resulting from a field-induced anisotropy. 

  The calculated curves are in fair agreement with the
experimental results for
this material.  In this context, it is worthy of mention that
the effect of
static disorder induced by the vacancies (absent spins) on the
shorter chain 
may become important at very low $T$ [13].  This effect is
absent in the above 
calculation. 

To account for disorder, we use the results in [16] where an
arbitrary distribution of impurities (which do not destroy the
integrability) parametrized by the real numbers $\theta_j$ at
the site
$j$ on the spin chain, are added to
a Heisenberg spin chain. This approach has been applied by Zvyagin [17] for 
the same purpose in the case of $Yb_{4}As_{3}$ recently.
The total free energy of the spin
chain with
impurities is $F = L^{-1}\sum_jf(\frac{\pi}{2}\theta_j)$ where
the sum
is taken over all the sites (for sites without impurities we
take
$\theta_j =0$), and $f(\theta_j)$ is given in (6). Now an
appropiate
random distribution of values 
$\theta_j$, $P(\theta_j) \propto \exp(-\pi (\lambda -
1)\vert\theta_j\vert)$
can be taken resulting in a power law 
divergence of the susceptibility at very low temperatures and
the
linear coefficient of the specific heat $\gamma$
\[
\langle \chi\rangle \propto \langle \gamma\rangle \sim
T^{\lambda -1},
\]
taken $\lambda \sim 0.26-0.42$ coincide with data on real
disordered
quasi 1D spin $\frac{1}{2}$ AF systems.  This is quite
consistent with the 
observed power-law divergence in $\langle \chi\rangle $.  It
would be 
interesting to see whether $\langle \gamma\rangle $ also shows
a similar 
divergence at very low $T$.            

 The specific heat shows the linear low-$T$ contribution.  A 
comparison of the curves for two values of $n$ shows that the
$\gamma$ 
coefficient of the linear term increases as $n$ increases
towards unity, providing a simple explanation for the larger
$\gamma$ value for $Yb_{4}P_{3}$ 
compared to $Yb_{4}As_{3}$.  The specific heat in an external
magnetic field
has been studied [4] by mapping the spin model in a staggered
transverse field
(induced by the uniform longitudinal field) onto a sine-Gordon
model, and 
accounts naturally for the field-induced opening up of the gap
in this material.
  Thus, the magnetic fluctuations are well understood in terms
of the 
excitation spectrum of the 1D Heisenberg model, as was inferred
previously [1-4].

  To understand the origin of the Fermi liquid characteristics
in transport 
, we observe that the small fraction of the localized spins in
the 
three elongated chains can be modeled by an Anderson-like model
with 
large $U$ (since the Yb holes in the other three chains do not
order).  The Hamiltonian is then given by,
\be
H = \sum_{k\sigma}C_{k\sigma}^{_\dag}C_{k\sigma} +
\sum_{ik\sigma}t_{k}(f_{i\sigma}^{\dag}C_{k\sigma}+h.c) +
\sum_{i}Un_{if\uparrow}n_{if\downarrow}
+ \sum_{i\sigma}(\mu n_{ic\sigma} + E_{f}n_{if\sigma})
\ee
where the $f_{i\sigma},f_{i\sigma}^{\dag}$ represent the $Yb$
4f fermions 
localized on the longer chains.  In this case, we use the
approach developed 
in [14] to compute the local spectral density and the
resistivity in $d=\infty$.
  In this case, our effective Anderson model yields a
correlated Fermi liquid
metallic state off half-filling (which is applicable to
$Yb_{4}As_{3}$, with
$n_{f} \simeq 0.1$).  The actual lattice coherence scale,
$T_{coh}$, can be
vastly different from the single-impurity Kondo scale, $T_{K}$
[14]; 
 The $c$-fermion selfenergy
 can be computed from the $d=\infty$ approach [15]; in the
low-$T$ regime, it 
takes on the simple form,
\be
Im \Sigma_{c}(\omega) =
\frac{(U)^{2}}{D^{3}}[\omega^{2}+\pi^{2}T^{2}]x(1-x)
\ee
where $D$ is the free bandwidth of the As ($2p$ band(s)) and
$x=n_{f}$.
This immediately tells us that the resistivity follows the
$T^{2}$ law 
characteristic of FL behavior.  The coefficient of the $T^{2}$
term is given by
$A=\frac{(U)^{2}}{D^{3}}$ and is large for $U>>D$.  
Notice that external pressure would increase $D$, reducing the
coefficient $A$, as indeed observed.

  One might wonder why the shorter chain that gives rise to the
observed 1D
magnetism does not seem to affect the transport properties.  To
see how this
can come about, recall that in the spin-charge separation
scenario, the holon
and spinon velocities are $v_{h} \simeq xt_{f}$ and $v_{s}
\simeq I_{eff}$.
These are much smaller than the Fermi velocity of the $2p$
carriers, as deduced
from the bandstructure calculation [6].  Given that
$v_{F}>>v_{h},v_{s}$, the
scattering process involving the $2p$ hole and a spinon or a
holon is unfavorable from the viewpoint of energetics (this
process would dominate if $v_{s}<v_{F}<v_{h}$).  Thus the heavy
fermion FL characteristics involves, in our picture,
the scattering of the light $2p$ holes off the (almost
localized on the time scale of the $p$ fermion hopping)
localized $f$ spins on the other three chains in
$Yb_{4}As_{3}$. In this context, we mention that it has been
proposed in [17]
that scattering of the As $p$ holes off the spinon excitations
of the squeezed
spin chain gives a resistivity going like $T^{2}$.  If this
were indeed true,
one would expect the $T$-dependence of the resistivity to
change upon the
application of an external magnetic field (which induces a spin
gap) in 
contradiction with experimental evidence.  The small magnitude
of the magnetoresistance 
 and the $\rho(T,H) \simeq A(H)T^{2}$ behavior
seen experimentally is thus in contradiction to the proposal in
[17], but is completely 
consistent with the mechanism proposed here.
  In any case, it is highly improbable that scattering of light
fermions off the spinons (holons) could yield such a large
$T^{2}$ term in the
resistivity.

  If the picture proposed above is correct, a more perfect
charge ordering of
the $Yb$ $f$ holes along chains should disfavor the onset of
heavy fermion FL 
formation.  This connection is indeed borne out by experiments
on $Yb_{4}P_{3}$ where a more perfect CO state is accompanied
by insulating behavior with a 
larger $\gamma$ co-efficient.  The more perfect CO state is a
signature of
decreased effective $f-d-p$ hybridization, a corresponding
decrease in the 
tendency to itinerance, and is consistent with the insulating
resistivity 
observed for this material. A more perfect CO state also
increases the 
number of spins on the shorter chain, and the higher
$\gamma$-coefficient
 is easily understood as arising from the 
spinon excitations of the squeezed Heisenberg chain, as shown
before. 

  In conclusion, we have proposed that the apparently strange
features in the
low-$T$ metallic state in $Yb_{4}As_{3}$, i.e, a heavy fermion
metal coexisting
with 1D quantum antiferromagnetism, are readily understood to
arise from a 
single observation:  the charge-order arising at higher $T$ is
not complete,
and a small fraction of the $S=1/2$ moments on $Yb$ reside on 
the three longer chains.  The 1D antiferromagnetism is
naturally understood 
in terms of the squeezed Heisenberg chain, while the heavy
fermion FL 
properties at low-$T$ result from the compensation of the $Yb$
moments on the
other three chains (the problem can be treated as
three-dimensional) by the
low density of carriers (coming from the As 2p bands). 

\bf{Acknowledgments}
\normalfont
J.R.R.M. wants to thank Prof. A. Kl\" umper for discussions and for
kindly making available his computer routines.
M. S. Laad wishes to thank Prof. P. Fulde for many discussions related to 
$Yb_{4}As_{3}$ and to Prof. M\"uller-Hartmann for a fruitful discussion.

\bf{References.}
\normalfont

[1] A. Ochiai {\it et al.}; J. Phys. Soc. Japan. {\bf 59}, 4129
(1990).

[2] M. Kohgi {\it et al.}; Phys. Rev. {\bf B56}, R11388 (1997).

[3] P. Fulde {\it et al.}; Europhys. Lett. {\bf 31}, 323
(1995).

[4] see, for example, S. Takagi {\it et al.}; Physica B {\bf
281-282}, 462
(2000) and references therein.

[5] M. Oshikawa {\it et al.}; J. Phys. Soc. Jpn. {\bf 68}, 3181
(1999).

[6] for the experimental evidence, see K. Iwasa {\it et al.};
Physica B
{\bf 281-282}, 460 (2000), and references therein.  For the
results of the
LDA calculation, see V. Antonov {\it et al.}, Phys. Rev. B {\bf 58}, 9752 (1998).

[7] P. Fulde {\it et al.}; Europhys. Lett. {\bf 31}, 323
(1995).

[8] Y. M. Li {\it et al.}; Phys. Rev. Lett. {\bf 78},3386
(1997).

[9] H. Shiba, Phys. Rev. B{\bf 6}, 3, 930 (1972).

[10] for a detailed exposition of these ideas, see, N. Andrei,
in "Lecture
Notes of the Trieste Workshop on Integrable Field Theories",
ICTP, Trieste
(1994).

[11] these ideas become formally exact as $U\rightarrow\infty$,
where the
total wave-function can be expressed as a direct product of a
spinless fermion
and a squeezed Heisenberg part.

[12] A. Kl\"umper {\it et al.}; Phys. Rev. Lett. {\bf 84}, 4701
(2000),  Eur. Phys. J. {\bf B 5}, 677 (1998).

[13] F. Steglich {\it et al.}, Z. Phys. B: Condens. Matter. {\bf103}, 235 (1997).  

[14] A. Georges {\it et al.}; Revs. Mod. Phys. {\bf 68}, 13
(1996).

[15] the c-fermion self-energy of this form follows from second
order 
perturbation theory, but in fact is already close to the actual
lattice 
self-energy in $d=\infty$. 

[16]A. Kl\"umper and A.~A. Zvyagin, Phys. Rev. Lett. {\bf 81},
4975
(1998);  A.~A. Zvyagin, Phys. Rev. B {\bf 62}, 6069 (2000).

[17] A.~A. Zvyagin, Phys. Rev. B {\bf 62}, 12175 (2000).

\bf{FIGURE CAPTIONS.}

% Figure 1
\begin{figure}
\epsfxsize=5in
\centerline{\epsfbox{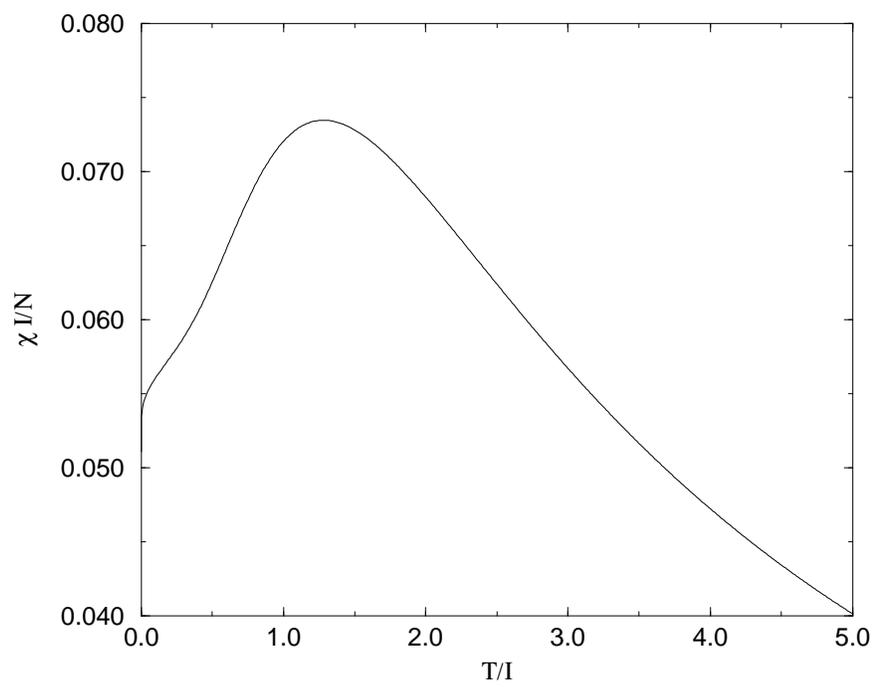}}
\vglue 0.1in % space between figure and caption
\caption{Magnetic susceptibility $\chi$ at low temperature $T$
for the
  spin $S = 1/2$ antiferromagnetic uniform Heisenberg chain of
length $N$. }
\label{Fig1}
\end{figure}

% Figure 2
\begin{figure}
\epsfxsize=5in
\centerline{\epsfbox{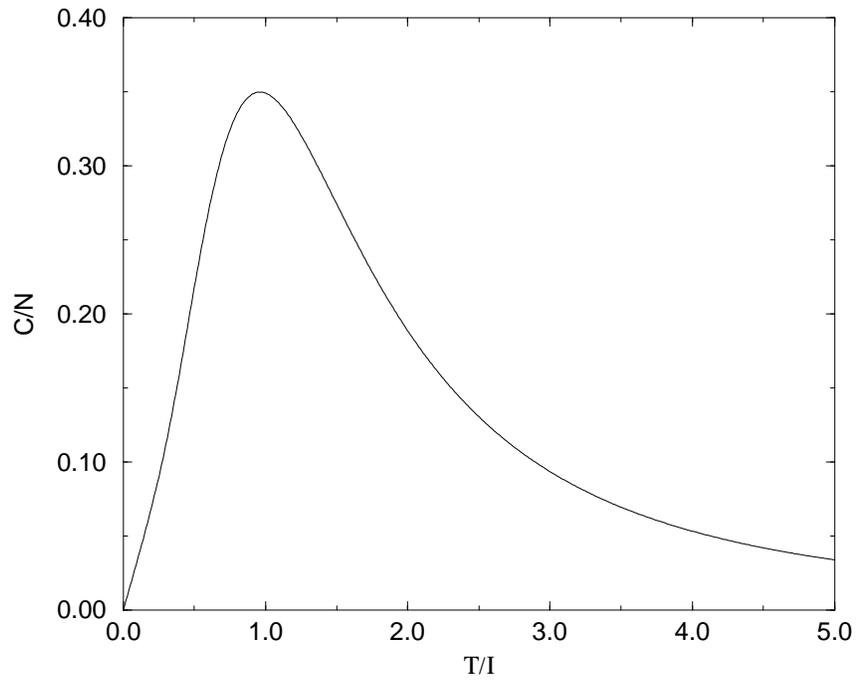}}
\vglue 0.1in
\caption{Electronic specific heat  $C$
  versus temperature $T$ for the $S = 1/2$ AF uniform
Heisenberg chain.}
\label{Fig2}
\end{figure}

\end{document}